\documentclass[12pt]{article}
\usepackage{amsmath, amssymb}
\usepackage{graphicx}
\usepackage{hyperref}
\usepackage{dirtytalk}
\usepackage[a4paper, total={7in, 10in}]{geometry}
\title{\textbf{How the Michelson and Morley Experiment Was Reinterpreted by Special Relativity} \footnote{  This is a revised and corrected translation of an article originally published in Spanish by the authors in 2005: “La reinterpretación radical del experimento de Michelson-Morley por la relatividad especial”. Scientiae Studia (Brasil), 3: 547-581.}}
\author{Alejandro Cassini \\ Universidad de Buenos Aires-CONICET Argentina \\ \href{alejandrocassini@gmail.com}{alejandrocassini@gmail.com} \and Leonardo Levinas \\ Universidad de Buenos Aires-CONICET Argentina \\ \href{leolevinas@gmail.com}{leolevinas@gmail.com}}
\date{}

\begin{document}

\maketitle

\begin{abstract}
We elucidate how different theoretical assumptions bring about radically different interpretations of the same experimental result. We do this by analyzing Einstein’s special relativity as it was originally formulated in 1905. Then we examine the theory’s relationship with the result of the 1887 Michelson-Morley experiment. We point out that in diverse historical contexts one and the same experiment can be thought of as providing different –often incompatible– conceptualizations of phenomena. This demonstrates why special relativity prevailed over its rival theories. Einstein’s theory made a new reinterpretation of the Michelson-Morley experiment possible by associating it with a novel phenomenon, namely the invariance of the speed of light, a phenomenon that was not the one originally investigated in that experiment. This leads us to an understanding of how this experiment could have been interpreted in a completely different historical context, such as Seventeenth-Century science when Earth’s orbital motion was still a questionable hypothesis.\vspace{0.4cm}

Pacs numbers: 03.30.+p \href{https://ufn.ru/en/pacs/03.30.+p/}{Special relativity}, 03.50.De \href{https://ufn.ru/en/pacs/03.50.De/}{Classical electromagnetism, Maxwell equations}, 07.05.Fb \href{https://ufn.ru/en/pacs/07.05.Fb/}{Design of experiments}, 01.65.+g \href{https://ufn.ru/en/pacs/01.65.+g/}{History of science}, 01.70.+w \href{https://ufn.ru/en/pacs/01.70.+w/}{Philosophy of science}\vspace{0.4cm}

 Historical context of the experiment. Assumptions and auxiliary hypotheses. Speed of light. Invariance of the speed of light. Earth’s motion.
\end{abstract}

\section*{Introduction}
The celebrated Michelson-Morley experiment (hereafter referred to as the $M-M$ experiment) conducted in 1887 is commonly associated with the decline of the theory of ether and the emergence of the theory of special relativity. A textbook on elementary particle physics, for example, provides the following description:

\begin{quote}
    This experiment provided clear proof that no such ether exists and that the speed of light is constant regardless of the motion of the source. (Coughlan, Dodd, \& Gripaios 2006, p. 10).
\end{quote}

Another text, in this case one about cosmology, states the following:
	\begin{quote}
	    In 1887, Albert Michelson and Edward Morley found that the speed of light is the same in 	all directions on  the Earth’s surface. (Harrison, 2000, p. 206).
	\end{quote}
	
 Even in one of the best advanced texts about relativity, we can read that:
 \begin{quote}
     Measurements first performed by Michelson (1881) showed complete lack  of the velocity  of light on  respecton its direction of  propagation. (Landau and Lifchitz, 1994, p. 3).
 \end{quote}

	Finally, a popular book written by an eminent specialist in relativity, presents the result of the experiment in these terms:
 \begin{quote}
     [...] In 1887 Michelson and Morley performed  their famous experiment designed to determine with high precision the change in the measured velocity of light due to the motion of an observer through the ether. They found no change wahtsoever. As a result of this  experiment, the ether theory  was overthrown, and the  principle of relativity was confirmed. (Wald, 1992, pp. 16-17).
 \end{quote}

	These quotations, selected almost arbitrarily, are just the tip of the iceberg. Each one offers typical retrospective interpretations of the result of the $M-M$ experiment in light of the theory of special relativity, a theory that was actually created eighteen years later. At the time this experiment was carried out, it was not interpreted in terms that could, in theory, have any direct relation to relativity; the very historical context of theoretical physics did not allow for this kind of interpretation. In the years immediately following the $M-M$ experiment, there was no inclination to conclude that the ether was non-existent, nor that the speed of light was constant even though the light source was in motion relative to the ether. Moreover, no one thought that the principle of relativity - the equivalence of all inertial frames of reference for the description of electromagnetic phenomena- would be confirmed. Nor did anyone think the hypothesis that the speed of light was invariant, that is, the same in any inertial frame of reference, would be confirmed.  What exactly did this experiment confirm or refute? This is a simple question, but its answer is incredibly complex and leaves no room for an unequivocal answer like the ones we just cited. We will try to shed light on the topic in the rest of this article. Examining the nature of assumptions in detail, whether explicit or implicit, that operate in any experiment and the role they play in the interpretation of the results, will all form essential elements of the analysis.
 
	The structure of this article is the following: In section \ref{section1} we present conceptual difficulties and incongruities that we consider to be the most important ones raised by the postulate that the existence of a luminiferous ether is an indispensable material medium  for propagating light waves before Michelson began his experiments in 1881. In section \ref{section2} we formulate, through a brief retrospective history, the most important issues brought up around the motion of light and bodies in a so-called luminiferous ether, and the dynamic state of the ether itself. In section \ref{section3} we describe the 1887 $M-M$ experiment and critically analyze some of its assumptions. In section \ref{section4} we discuss two completely different interpretations of the experiment. We discuss that of Michelson and Morley themselves and that of Lorentz, and we show that, in fact, they lead to two different conceptions of not only the character of motion but also of the nature of the bodies, and this touches Newtonian mechanics directly. In section \ref{section5} we present the postulates of the theory of special relativity in such a way that its 1905 formulation can be understood to be independent of the result of the $M-M$ experiment and of any of the interpretations that were available at the time. We stress the fact that even if in its original formulation, it does not rely on the existence of the ether, special relativity does not necessarily imply that the ether does not exist. Moreover, Einstein reintroduces the ether within the context of general relativity. In section \ref{section6} we analyze the interpretations of the $M-M$ experiment after 1905 and discuss how a new interpretation places the experiment at the front and center, making it decisive regarding a phenomenon (the invariance of the speed of light) that was not the original research focus. We emphasize the different roles that this experiment played in the contexts of discovery and justification of special relativity. In the conclusion (section \ref{section7}) we determine, by analyzing some noteworthy examples, the importance of historical context and the corresponding assumptions that operated while working through the problem of measuring the speed of light. Along the way, our sights are set on establishing reasons why special relativity, despite being such an unintuitive theory, became the predominant theory over rival ones.

\section{Light and ether} \label{section1}
During the first half of the 19th century, the wave theory of light seemed to have been properly confirmed by the diffraction and interference experiments carried out by Young, Fresnel, and others. The prevailing interpretation of these experiments was that the wave theory of light explained the phenomena that the corpuscular theories could not. According to measurements taken by Fizeau in 1849, the light was conceived as an oscillation with a propagation speed on the order of 300,000 km/s. That said,  mechanical waves, such as sound, ocean, or heat waves, consist of vibration in a material medium such as air or water. However, light, unlike sound, could propagate through interstellar space, evidently a vacuum. One way to solve this problem was to assume that the space was not a vacuum but rather filled by a subtle medium called luminiferous ether. Apparently, luminiferous ether was thought of as a type of materialization of the absolute space that Newton required: continuous, homogeneous, and isotropic. The corpuscles or particles were believed to be mutually impenetrable, meaning that one particle could not pass through another. In contrast, it was believed that the ether, as a different type of material medium, could penetrate these particles. Thus, the ether seemed to have both characteristics of a vacuum and as well as those of matter. Like a vacuum, it had the absence of mass, the lack of resistance against motion, and no density at all. Like matter, it had elasticity and the ability to oscillate or vibrate. 

Now, we should note that sound is propagated in a microscopically discontinuous medium, which is different from the case of light in a \say{vacuum} that should be filled by a truly continuous ether -if the ether were not continuous, even if the discontinuity were minor, such ether would not be necessary as a medium for mechanical waves. That is, if light could not propagate through the vacuum in interstellar space, then, neither could it do so through the empty space between the atoms of matter (for example, between water molecules). It seems reasonable then to identify the ether with space itself. From this point of view, it even becomes more difficult to explain how other mechanical waves like sound can propagate through a discontinuous medium like the air. 

Maxwell, who was grounded in Faraday's qualitative ideas, developed a mathematical theory in the 1860s that unified all the known phenomena of electricity and magnetism  \footnote{In this paragraph and the following, we offer a rough sketch of the context of optics and electromagnetism that surrounded Michelson and Morley's experiment. The work by Born (1962), chapters 4 and 5, contains a detailed exposition of the optical and electrodynamic theories of the 19th century. Balibar (1992) comments on Einstein's perspective about those theories.}.  Maxwell also postulated the existence of some type of ether as a medium for electromagnetic waves. In Maxwell's equations, a constant $c$ represents the speed of propagation of these transverse waves-waves whose direction of propagation is perpendicular to the plane of oscillation of the electric and magnetic fields. As early as 1817 Fresnel had proposed a hypothesis that light consisted of transverse waves, and since then this hypothesis has been accepted as the only possible explanation for the phenomenon of light polarization. On the other hand, the calculation of the constant $c$ based on the equations of Maxwell demonstrated that it had a value almost identical to that of the speed of light, as had been determined empirically. Based on these coincidences, Maxwell surmised that the light was nothing more than a special type of electromagnetic wave.

	Maxwell's conjecture was later confirmed by the experiments with radio waves conducted by Hertz in 1888. Hertz proved that this type of waves produced reflection, refraction, interference, and polarization just like waves of light. Additionally, by knowing the frequency of radio waves, he was able to calculate their length and, consequently, their speed of propagation, which turned out to be equal to the speed of light. Considering these facts, Hertz believed that Maxwell's conjecture was correct, and, from that moment, the community of physicists accepted that light was an electromagnetic wave. Thus, optics and electromagnetism were unified and the luminiferous and electromagnetic ether became two names for the same entity.
 
	Even if the theory of the electromagnetic field required the existence of the ether as a continuous medium in which waves were propagated, it did not determine its mechanical properties. The scientific community never reached a consensus on this point. Some physicists conceived the ether as a perfect fluid, while others imagined an elastic solid. Some thought that it was immobile with respect to fixed stars, but others suggested that it was dragged, totally or partially, by moving bodies. All these hypotheses presented diverse conceptual and empirical anomalies that were never successfully resolved. Soon, it seemed clear that it was difficult to imagine a mechanical model for the ether, even though prominent physicists like Maxwell and Kelvin tried. If the ether were a material substance, undoubtedly it ought to possess very different properties from those of ordinary matter. For example, to transmit transverse waves of extraordinarily high velocities, like light, the ether would need to be a completely elastic solid. Yet, it should also have, as we have seen, a zero or almost zero density so as not to resist the movement of bodies and light itself. On the other hand, just like we have seen, it should be a solid that completely penetrates ordinary matter thus allowing the propagation of light in transparent material media such as air, water, or glass. In short, it was assumed that the light-ether interaction was either null or negligible at most. This should explain the fact that light waves apparently did not dissipate when propagating in the ether, unlike what happened with other mechanical waves in relation to their propagation medium. On the other hand, in 1850 Foucault verified that the speed of light in water is less than in air. This indicates some kind of interaction with the material medium. The concept was formally expressed in the equation corresponding to the evolution of the wave. Specifically, some characteristic quantity of the medium (such as the refractive index) was part of the solution of the equation. 
 
In Maxwell's equations, $c$ represents the speed of propagation of the electromagnetic waves through the vacuum. In Maxwell's theory, the speed of light that is emitted is constant, meaning, the same in every direction regardless of the movement of the source. This fact is shared by all wave phenomena. For these reasons, it seemed natural to think that the ether provided the frame of reference in which velocity $c$ was defined. The speed of light should be constant only in this frame of reference. In any other frame of reference that is moving at a constant velocity relative to the ether, the speed of light ought to vary depending on the direction in which it is measured. If a frame of reference moves at a velocity $v$ relative to the ether, the speed of a beam of light with respect to that frame of reference would be $c + v$ if the frame of reference were moving along the same path but in the opposite direction of the beam of light. Yet, it would be $c - v$ if were traveling along the same path and in the same direction. When the frame of reference is moving along a different path with respect to the beam of light, its velocity should have an intermediate value between $c + v$ and $c - v$. These are nothing more than consequences of the classical transformation of velocities incorporated in Newtonian mechanics. It then follows that velocity is never invariant with respect to an inertial referential change. Naturally, this law was assumed to be valid for electromagnetic phenomena as well.
	The existence of the ether as a privileged frame of reference in which the speed of light was constant, makes it possible, in theory, to measure the absolute velocity at which a body moves. Maxwell sustained that an electromagnetic experiment would be capable of determining that speed if it were sensitive enough to measure quantities on the order of $v^2/c^2$ (where $v$ is the velocity of the body with respect to the ether). In 1878 Maxwell himself suggested that the only possible experiment on Earth would consist in measuring the time that light takes to travel towards a given point and back. It would travel a certain distance along a path that is parallel to the orbital motion of the Earth (that is of the order of $1/10,000 c$). The time calculated should differ when traveling in the same direction as the Earth’s motion in comparison to the opposite direction. The difference between the two times would make it possible to calculate the speed of the Earth with respect to the ether. Maxwell believed that no technically conceivable experiment would have enough sensitivity to measure this difference. Nonetheless, the experiment was physically possible.

\section{The problem of motion with respect to the ether} \label{section2}

The problem of the motion of the bodies through the luminiferous ether laid out severe difficulties for the wave theory of light during the 19th century. Was the ether immobile and the bodies moved through it freely? Or did the bodies drag all or part of the ether with them as they moved? In this article, we do not intend to even outline the complex history of the different responses that physicists gave to this question. Nonetheless, we think it necessary to point out some facts about Michelson and Morley's celebrated experiment and the theoretical hypothesis that underpinned their optics theory.

	In 1728 Bradley attempted to measure the annual parallax of the stars. The expected outcome was a record of an elliptical movement, parallel and opposite to the terrestrial orbital movement. The amplitude of the elliptical movement was expected to be dependent on the distance of the star from Earth. However, the luminous phenomenon that Bradley observed indicated that the star, although it \say{moved} in an elliptical fashion, did so perpendicular to the terrestrial trajectory. This \say{aberration} effect was compatible with a corpuscular conception of light, such as the one held by Bradley himself, whenever the Earth was in motion since the result could be explained by combining the terrestrial orbital velocity with that of light. So, a star located  perpendicular to the movement of the instrument could be seen in the center of the visual field, the telescope had to have an angular inclination $\alpha$ given by $\alpha = \arctan(v/c)$, $v$ being the orbital velocity of the Earth. The angle $\alpha$ changed during the year since the velocity $v$ was not constant. In this scheme, it was possible to take Galileo's law of composition of velocities to calculate the velocity of light in the telescope from any reference system for stars located at any position with respect to the Earth. It should be noted that  the calculation, in fact, took $v < c$. 
 
	Now then, the wave theory faced serious difficulties for more than a century when trying to account for this phenomenon. Basically, it was challenging because the angle of aberration should not depend on the speed of the telescope with respect to the star (where the light source was), but rather on its velocity relative to the propagation medium of light, the ether. With the detection of the aberration of light, a traumatic story began, in which it was attempted  to reconcile this phenomenon with the belief that light had wave characteristics. It is not our intention -as already stated- to describe this history exhaustively, but rather to point out a few important episodes with the aim of bringing attention to the context in which the $M-M$ experiment was originally interpreted. 

In 1804, Young argued that the phenomenon of stellar aberration was explainable within the framework of the wave theory of light if it was assumed that the orbital motion of the Earth did not affect the absolute immobility of the ether (Young, 1804). In 1818, Fresnel proposed a different explication of the aberration phenomenon. He postulated that the substances transferred a fraction of their movement to the ether each contained inside of itself. Fresnel based his hypothesis on results obtained by Arago (at that time a corpuscularian philosopher!), who in 1810 had shown that the aberration was not affected by refraction in a prism (Arago, 1810). The angle of aberration $\alpha'$ (different from that of the vacuum) was the same whether the light of a star was observed while facing the direction of the Earth's motion or the opposite direction. In every refraction, beams of light change their speed  -from $c$ in a vacuum to $c/n$ in a refractive medium-  ($n$ being the refractive index of the medium). Arago's experiment was evidently incompatible with Young's hypothesis. If the ether was immobile, the velocity of light in the direction of the movement of the Earth should be different from its velocity in the opposite direction. To provide an explanation within the framework of the wave theory of light for Arago's null result, possibly the first experimental result against ether theories, \footnote{See Ferraro and Sforza (2005) for a detailed explanation of this hypothesis.}  Fresnel proposed that there must be a drag factor whose value was $f = 1 - n^{-2}$. Consequently, the speed of light in a refractive medium was equal to $c/n + fv$ (where $v$ is the velocity of the medium with respect to the ether). This hypothesis, in that moment purely \textit{ad hoc}, compensated for the \textit{partial} drag of the ether and the aberration of the light, in such a way that Snell's law holds in a system that is moving together with the Earth at an order of $v/c$ (Fresnel, 1818). 

However, in 1845, Stokes denied that this had happened and assumed that the ether was completely dragged by the Earth and the bodies located on its surface (Stokes, 1845). He postulated that the ether was a viscous fluid that produced friction with the Earth in movement. The friction of the Earth dragged it, creating a series of layers, in such a way that the ether was dragged completely at the level of the surface until at a certain altitude it reached a complete immobility. This hypothesis immediately explained Arago's null result since the prism was always found at rest with respect to the ether, and thus the speed of light should be the same along any path or in any direction. Nonetheless, to explain the aberration, Stokes had to assume that the dragged ether did not have any rotatory movement, an extremely implausible hypothesis considering mechanical fluids. Sometime later, in 1886, Lorentz pointed out that Stokes's ether must have a field of velocities such that it would require inadmissible boundary conditions (Lorentz, 1886).

	For his part, in 1851 Fizeau confirmed Fresnel's prediction within a 15\% margin of error, through an optical experiment, independent of Arago's, which consisted of measuring the speed of light in water currents circulating in opposite directions. Fizeau also found that the entrainment of ether by air was negligible and, for any optical experiment, could be considered identical to that of a vacuum (Fizeau, 1851). Another favorable test for Fresnel's theory was obtained by Airy in 1871 when he showed that the angle of aberration did not change in a telescope filled with water, as deduced from the partial ether drag coefficient hypothesis (Airy, 1871). These two experiments provided independent confirmation of the partial ether drag hypothesis and seemed to remove the \textit{ad hoc} character it had when Fresnel introduced it to explain the result of Arago's experiment.
 
            In 1874 Mascart culminated a series of experiments in $v/c$ order  -using both terrestrial and sunlight light sources - with negative results regarding the detection of differences in a system moving together with the Earth with respect to a system supposedly at rest. He concluded that \say{these phenomena do not give us a way to appreciate the absolute movement of a body and that the relative movements are the only ones that we can reach} (Mascart, 1874, p. 420). Fresnel's interpretation of partial drag omitted the fact that each frequency of light had a different refractive index, so the ether drag must depend not only on the medium but also on the light frequency (color). Mascart also recognized this fact in a double refraction experiment. Given that the refractive index was different for the two beams, it followed that in Fresnel's model, different quantities of ether were dragged for each of the two beams \footnote{See Janssen and Stachel (2004), p. 15. In this article, a more extensive exposition of the problem of the dynamic state of ether can be found.}.
            
 	This was the context in which the experiments carried out by Michelson starting in 1881 were planned. Michelson, from novel and independent experiments, tried to determine the dynamic state of the ether based on three fundamental assumptions: the wave nature of light, the existence of a luminiferous ether, and the movement of the Earth.
  
\section{	The Michelson and Morley Experiment}\label{section3}

Michelson, by accepting Maxwell's challenge, designed a device that was sensitive enough to detect effects of the order of $v^2/c^2$. The aim of the experiment was to measure the relative velocity of the Earth with respect to the luminiferous ether using a device that allowed light to travel through the $air$. For that reason, they could disregard the Fresnel dragging coefficient. Given that the experimental devices designed and employed by Michelson have been described on many occasions, we can be brief as we touch on the subject. 

  In his 1881 experiment, a beam of light is split by a beam splitter into two perpendicular beams. One beam is directed towards a mirror, reflected to the starting point, while the other beam is directed towards a second mirror, reflected to the same starting point. At their shared starting point, interference fringes are formed due to small differences in the respective distances to the mirrors and their different inclinations. To create sharper interference fringes, a precision screw is used to adjust the distance or inclination of the mirrors. If the apparatus is rotated slowly up to 90°, the wave theory of light -together with classical mechanics- predicts the interference fringes will be displaced due to the change of direction of the instrument with respect to the direction of the Earth’s orbital motion. 
  
Michelson conducted this experiment without observing the predicted changes in the interference fringes and came to the following conclusion: \say{The interpretation of these results is that there is no displacement of the interference bands. The result of the hypothesis of a stationary ether is thus shown to be incorrect, and the necessary conclusion follows that the hypothesis is erroneous} (Michelson, 1881, p. 128). This experiment was inconclusive basically due to a calculation error in the combination of velocities in the arm of the instrument oriented perpendicular to the movement of the Earth. That same year, when Michelson traveled to Paris to present his results, Potier pointed out the difficulty, although Potier's own calculation also turned out to be wrong. In 1882, Michelson published a corrected version of the calculations from his experiment (Michelson, 1882). In 1886, Lorentz made a detailed analysis of Michelson's experiment and independently performed the correct calculations (Lorentz, 1886) \footnote{Michelson's error consisted of not combining the Earth's orbital velocity with the speed of light in the vertical arm of the instrument. Potier suggested that the corrected calculation implied that there was no displacement of the interference fringes. Michelson, on the other hand, came to the conclusion that the displacement of the spectral lines should be reduced by half, that is, from 0.08 to 0.04 of the width of an interference fringe, very close to the observational error of 0.02. In his 1886 paper, Lorentz also pointed out that if a possible ether drag on the Earth's surface was taken into account, according to the Fresnel dragging coefficient, the ether wind speed should be halved, and, therefore, also the displacement of the interference fringes. Michelson knew about Lorentz's work through Lord Rayleigh, but by then he already had the right calculations. The reply letter to Rayleigh, March 6, 1887, shows that until that date, a month before beginning the experiment with Morley, Michelson was unaware of Lorentz's article (see Shankland, 1964, p. 29), an article that he then quoted twice in his written work with Morley published in November 1887 (Michelson and Morley, 1887, pp. 334 and 335).}. 

In the same year, Michelson and Morley made a more precise measurement of the Fresnel ether dragging coefficient. The experiment confirmed his predictions and encouraged them to assume that Fresnel's theory was the correct explanation of the phenomenon of stellar aberration (Michelson and Morley, 1986).  \footnote{  The ether dragging coefficient for water predicted by Fresnel was 0.438. The value measured by Fizeau in his 1851 experiment was $5.0 \pm 0.1$, while the value measured by Michelson and Morley in 1886 was $0.434 \pm 0.03$. }

To show that Fresnel's explanation was satisfactory, some measurements had to be made in a medium whose coefficient was negligible, this is, the usual ether drag did not occur. (In this case $f = 1 - n^{-2}$, such that $n \approx 1$ and $f \approx 0$).  Consequently, in this experiment, a wind ether ought to have been produced by the movement of the Earth. Thus, in 1887, Michelson and Morley conducted a new experiment with an improved device that multiplied the distance traveled by the light more than nine times by using a multiple reflective system. They corrected the 1881 calculations by considering \say{the effect of the motion of the earth through the ether on the path of the ray at right angles to this motion} (Michelson and Morley, 1887, p. 334). If the distance traveled by the light in each arm of the apparatus is equal to $L$ (ignoring the difference in their lengths, of the order of the wavelength of the light used), if the apparatus were at rest with respect to the ether, the time spent by light in a round trip should be equal to $2L/c$ in each arm. If the orbital speed of the Earth $v$ is taken into account, those times must be different when one of the two arms is in a direction parallel to this movement and the other in a perpendicular direction \footnote{The corresponding calculations, which we omit here, can be found in any textbook on special relativity (for example, French, 1968, Ch. 2, or Resnick, 1968, Ch. 1). Two easily accessible works are those of Mills (1994), pp. 31-38; and Sartori (1996), pp. 29-39. The last one mentioned is one of the few books that offers complete calculations of the time taken by light in its round trip in each arm of the instrument in two different frames of reference: that of the ether (where it is assumed that the speed of the light is constant but the distance traveled by light in two mutually perpendicular directions is different) and that of the Earth (where the distance traveled by light in two perpendicular directions is the same but the speed of light is not constant but instead is combined with the orbital speed of the Earth). Both calculations obviously give the same result since, in classical mechanics time is invariant with respect to an inertial referential change. Jaffe (1960) and Shankland (1964) contain valuable historical information and should also be mentioned.}.  In the first case, the calculations demonstrated that the total time $T_1$ was equal to $\frac{2L}{ c  (1-v^2/c^2)}$, while in the second case, time $T_2$ was equal to $\frac{2L}{ c  (1-v^2/c^2)^{1/2}}$.  Given that $\frac{v^{2}}{c^2}<1$ , it turns out that $\sqrt{1-\frac{v^{2}}{c^2}} > 1-\frac{v^{2}}{c^2} $  , and, consequently, that $T_1 > T_2$. The difference between the two times is then 
$$T_1 - T_2 = \frac{2L}{c (1-\frac{v^{2}}{c^2})} - \frac{2L}{c \sqrt{1-\frac{v^{2}}{c^2}}} =\frac{2L}{c} \left( \frac{1}{1-\frac{v^{2}}{c^2}}- \frac{1}{\sqrt{1-\frac{v^{2}}{c^2}}} \right). $$ And this, neglecting powers higher than second, is approximately equal to  $\frac{L}{c} (\frac{v^{2}}{c^2})$, which is a very small number.   \footnote{The difference between the two paths of the rays is, in turn, $L \frac{v^{2}}{c^2}$ , again, a small number. In the 1887 experiment where $L$ increased by a factor of almost 10 with respect to the 1881 experiment, the shift of the interference fringes increased from 0.04 to 0.4 of the width of a fringe. This offset was clearly observable since the difference was far above the margin of observational error, 0.01}

The deduction of these two relationships was only possible based on a large set of presupposed hypotheses. It would be difficult to list them all, so without claiming completeness, we will point out the following: 1. The Earth moves around the Sun at an orbital velocity of the order of 30 km/s. 2. The ether is approximately at rest relative to the Sun. 3. The ether is practically not dragged at all by the Earth, and, as a result, it is in relative movement with respect to it. 4. The speed of light is constant with respect to the ether. 5. The speed of light is not constant in the frame of reference of the Earth in motion but rather composed with the orbital velocity of the Earth, according to the classical or Galilean transformation of velocities. 6. The light is a wave phenomenon that reveals itself in interference fringes produced by beams of light that converge on the mirror. 7. The length of the arms of the instrument remains unaltered regardless of the direction of their motion, specifically, rotating the device does not affect its length. There may be other presupposed hypotheses that remain implicit and are, therefore, difficult to identify and bear weight in the interpretation that we cannot specify.  \footnote{The experiment, like any other, also makes idealizations. The Earth is not an inertial system. It moves in
an almost circular movement and is in rotation. Nonetheless, the effects of acceleration due to the change
in the direction of its velocity can be ignored because the time the light takes in the experiment is so short.
On the other hand, the reflection process of light in moving mirrors is also idealized, in particular, the
change in the angle of reflection is ignored. However, it is not evident that the effects of reflection are
irrelevant to the results of the experiment.}

Michelson and Morley admitted that it was possible that the Sun, and with it the entire Solar System, were in movement with respect to the ether. If this were so, the velocity of the Earth with respect to the ether was assuredly greater than its orbital velocity. The value of the latter with respect to the ether was a minimum that they expected to measure. Given that they were prepared to find a value that was much higher, they would have attributed an eventual excess of velocity to the movement of the Solar System. For this reason, the null result, that is $v = 0$, was entirely $unexpected$. In fact, it was a result that could not be accepted within the context of physics at the end of the 19th century since it would have meant that the Earth was at rest with respect to the cosmic ether. The null result of the experiment had to be explained some other way. \footnote{The result, in fact, was not exactly null. The data of the $M-M$ experiment exhibited a certain systematic
tendency. A recent statistical analysis demonstrates, however, that such data were compatible with the
null result and did not indicate that $v$ was different from zero (see Handschy, 1982).}

To understand the situation in which Michelson and Morley found themselves, we must make a few further clarifications about the 1887 experiment. Primarily, the instruments did not measure the velocity of light in different directions, nor the time taken by light in its journey to the mirror and back along each arm of the device. The only thing that was directly observable were the interference fringes produced by the beams of light when they appeared on their return trips. Using a graduated microscope with a reticle, it was also possible to observe the shifting of the interference fringes with a precision of one-hundredth of a fringe. Knowing the wavelength of the light used, the calculations predicted a maximum shift of four-hundredths of a fringe when the entire apparatus was rotated, which was within the range of resolution of the instruments used. If we call $N$ the number of interference fringes that pass through a reticle as the spectrum of fringes shifts, the null result of the experiment can be expressed, in terms of observable quantities, as $\Delta N = 0$. This result contradicts one of the basic assumptions of the experiment according to which $v \neq 0$, that is, that the Earth moves with respect to the stationary ether. The problem passed on by Michelson and Morley to the scientific community of their time consisted in explaining the null result ($\Delta N = 0$), without postulating that $v = 0$. The historical context of the experiment did not allow for interpreting it as proof that the Earth was immobile with respect to the ether. In a different context, like in the 17th century for example, the most natural and evident interpretation of the null result would have been precisely that $v = 0$. So, this class of experiment would have been taken as direct proof that the Earth did not move in space and, possibly, would have significantly delayed the general acceptance of the Copernican theory. At the conclusion of this article, we will come back to this point.

When an experiment produces a result that is incompatible with the expected predictions, detecting the source of the error is always a complex issue. More than a century ago, Duhem argued that in any experiment in physics it is never a single hypothesis or theory that is tested but rather a group of hypotheses or theories (Duhem, 1914, p. 278) \footnote{This thesis by Duhem already appears in an early article (Duhem, 1894), whose content is reproduced
with almost no change in the first and second editions, 1906 and 1914 respectively, of his fundamental
epistemological book (Duhem, 1914). Epistemological holism is commonly known in the contemporary
philosophy of science under the name of the \say{Duhem-Quine thesis}. This name is not particularly
appropriate because there are important differences between the theses of each of these two philosophers.
For example, while Duhem limits the knowledge presupposed in an experiment in physics to theories in
the domain of this science, Quine also includes considerable portions of logic, mathematics, and non-
scientific common-sense knowledge (see in this regard Gillies, 1993).}.  In principle, all the assumptions of the experiment are subject to revision. The problem is that the experimenter is not usually aware of all the assumptions of his own experiment. Any experiment has a considerable amount of assumed auxiliary hypotheses and theories. As Quine (1992, p. 17) pointed out, no scientist even tries to make a complete list of all the assumptions necessary to derive a certain testable prediction from a theory. This would probably be a never-ending task, since each assumption may in turn have other assumptions, and so on. Consequently, no experimenter nor any interpreter of the experimental results at a given historical moment knows the totality of the assumptions of the experiment that he or she performs or interprets. The review of an experiment, then, sometimes consists in finding a hidden assumption that is doubtful or directly unacceptable considering current knowledge.

	The $M-M$ experiment was conducted over just three days: July 8, 9, and 11, 1887. As the authors themselves indicated towards the end of their article, there was the possibility that the orbital motion of the Earth was compensated, almost exactly, by an eventual motion of the entire Solar System in the opposite direction. Though this coincidence was improbable, the best way to exclude it consisted in conducting the experiment six months later when the velocity of the Earth would change its direction. Specifically, in 1879 Maxwell had pointed out the need to verify that the Solar System moved or not with respect to the ether, employing a method analogous to that used by Römer with the eclipses of Jupiter’s satellites (that is, when Jupiter travelled in the same direction as the movement of the Solar System and in the opposite direction).  
 
	Another problem that the $M-M$ experiment brought up was that of the possible existence of another type of ether drag. In fact, it was possible that the walls of the laboratory where the instrument was enclosed dragged the ether in such a way that it was at rest with respect to the light source that was being used. Thus, it was suggested that the experiment should be repeated in the open air, preferably at a high place such as the top of a mountain, where, presumably, there would be no ether drag. There was also a similar problem with the opaque bodies, that would not allow ether to pass through. Given that Michelson and Morley's tools were isolated by a wooden covering, that was not transparent, it was possible that the ether was dragged by this covering.
 
	The third difficulty consisted in using a source of light on Earth and not, for example, starlight that could come from bodies at rest with respect to the ether. In principle, an experiment that used outside light could give a positive result. It is interesting to consider the first experiences related to this problem. Römer (in 1676) and Bradley (in 1728) conducted experiments using outside light, but they were not aiming at measuring the velocity of the Earth with respect to the ether. They presupposed that the Earth moved and sought to measure the speed of light or the parallax of the stars.
 
	Michelson and Morley did not repeat the experiment in such a way that they could respond to each of these possible objections. Nonetheless, the same type of experiment was conducted for many years at different altitudes, using sunlight or starlight, and even in an isolated environment surrounded by crystal, where the ether (of the laboratory) would not be dragged given that crystal was transparent to light. All these experiments gave negative results under the same type of interpretation offered by Michelson and Morley, that is, within the context of the ether theory. On no occasion was detected the existence of an \say{ether wind} of a magnitude even close to what had been predicted. The few positive results, such as Miller's in 1924-25, could be attributed to experimental error or disturbing influences, such as excessive temperature or the vibrations of the instruments. \footnote{Miller claimed that he had measured an 8 km/s ether wind, which was well within the scope of $M-M$'s
original experiment and subsequent ones. In 1933 Miller still considered his positive results to be due to a
genuine ether current and believed that experiments that produced null results did not reproduce all the
conditions of his own experiments (Miller, 1933). The issue was only clarified in 1955 when Shankland
and his associates made a detailed statistical study of Miller's data and concluded that these could be
explained by temperature variations in his instruments (Shankland \textit{et al.}, 1955). This had been Einstein's
own conjecture, on December 25, 1925, when Miller shared the results of his experiments with him (see
Holton, 1969).} 

\section{The Interpretation of the experiment before 1905} \label{section4}

Michelson and Morley never presented their experiment as crucial between the Fresnel and Stokes hypotheses regarding the dragging of the ether by matter, as is often argued (for instance, by Janssen and Stachel, 2004, p. 12). At the beginning of their 1887 paper, they indicated that their aim was to experimentally test Fresnel's hypothesis that the ether is at rest except within transparent media (Michelson and Morley, 1887, p. 334). However, in the conclusion they expressed the negative result in the following way: 
\begin{quote}

It appears from all that precedes, reasonably certain that if there be any relative motion between the earth and the luminiferous ether, it must be small; quite small enough entirely to refute Fresnel’s explanation of aberration. Stokes has given a theory of aberration which assumes the ether at the earth’s surface to be at rest with regard to the latter, and only requires in addition that the relative velocity has a potential; but Lorentz shows that these conditions are incompatible. Lorentz then proposes a modification which combines some ideas of Stokes and Fresnel, and assumes the existence of a potential, together with Fresnel’s coefficient. If now it were legitimate to conclude from the present work that the ether is at rest with regard to the earth’s surface, according to Lorentz there could not be a velocity potential, and his own theory also fails. (Michelson and Morley, 1887, p. 341)
\end{quote}

 Note the cautious and ambiguous tone of this conclusion that is purely negative. Generally, when an experiment does not offer the expected results, its designers discuss some of the assumptions included in the interpretation. Those that are reviewed first are always the ones that are considered the least obvious or least firmly established. Clearly, in the case of measuring the speed of light, an unshakable assumption was that the Earth moved around the Sun. Michelson and Morley were also unwilling to review the fact that light was a wave phenomenon that required the existence of the ether. When connecting the hypothesis of terrestrial movement with the previous conclusion, what seemed to prevail was that, on its surface, the Earth dragged the ether; therefore, Michelson and Morley considered the hypothesis that the ether was not dragged, to be disproved. The ether drag hypothesis implied that drag must be less in places that were further from the surface of the Earth. Consequently, the result of the same experiment should be different if it were conducted on top of a mountain. 
 
The conclusion, obviously, was not a necessary deduction; it was only one of many possibilities. From a logical point of view, it is possible to consider that the experiment disproved one of the assumptions that we mentioned before or any group of them. As is well known, FitzGerald in 1889 and independently Lorentz in 1892 proposed a different interpretation. They argued that the experiment refuted the hypothesis that the length of the arms of the instrument remained unchanged when it was in motion relative to the ether, a tacit assumption of the $M-M$ experiment. They then formulated the hypothesis that the length of rigid bodies that move with respect to the ether is not invariant, but rather contracts in the direction of motion by a factor equal to $(1-v^2/c^2)^{1/2}$, such that $L_v = L (1-v^2/c^2)^{1/2}$ (where $L_v$ is the length of a body that moves at a velocity $v$ with respect to the ether, and L is its length when it is at rest with respect to the ether). Accordingly, in the $M-M$ experiment, the length of the arm aligned with the motion of the Earth is shortened, while the length of the arm perpendicular to the motion remains unchanged. This is an absolute contraction of length, that is, a contraction with respect to the ether. This explains the null result of the experiment because the times of travel of the light in each arm of the instrument are now equal to $\frac{2L}{ c(1-v^2/c^2)^{1/2}}$. When the instrument is rotated, the arm lengths are offset in such a way that the light travel times always remain the same. For example, when both arms are at 45° to the direction of Earth's motion, each experiences a contraction equal to $1/2(1-v^2/c^2)^{1/2}$. This would explain why the interference fringes are always stable, without any observable shift.

The contraction hypothesis is usually considered to be purely \textit{ad hoc} since it was proposed solely to accommodate the result of the $M-M$ experiment in order to save the quiescent ether hypothesis from refutation. This affirmation is debatable and depends on what is understood by the concept of an \textit{ad hoc} hypothesis. \footnote{Zahar (1989), pp. 47-52, examines various ways in which the Lorentz contraction hypothesis could be
considered \textit{ad hoc}. He points out that Lorentz in 1892 deduced this hypothesis from another more general
hypothesis about the behavior of molecular forces acting on bodies in motion. He argues that the
hypothesis of contraction was not conceived exclusively with the aim of explaining the result of the $M-M$
experiment, and, in that sense, it was not an \textit{ad hoc} hypothesis. However, Lorentz undoubtedly knew
about the $M-M$ experiment and took it into account when formulating his electrodynamics of moving
bodies. On the other hand, in the brief note by FitzGerald (1889), published three years before Lorentz's
work, the contraction hypothesis was explicitly introduced to explain the result of the $M-M$ experiment,
and was therefore \textit{ad hoc} also in the sense pointed out by Zahar.} The truth is that the Lorentz-FitzGerald contraction is a hypothesis that is, in principle, arbitrary. In effect, there are many other hypotheses of the same type that could also account for the null result. For example, there is one that states that the arm oriented perpendicular to the direction of the Earth motion expands according to the factor $(1-v^2/c^2)^{1/2}$, such that $L’_v = L / (1-v^2/c^2)^{1/2}$ (where $L’_v$ is the length of a body that moves with a velocity $v$ with respect to the ether but it is perpendicular to the direction of the motion). This hypothesis is not empirically equivalent to the previous one since it predicts the times the light takes to travel to the mirror and back are equal to $\frac{2L}{ c(1-v^2/c^2)^{1/2}}$. Both hypotheses, the contraction, and the expansion hypotheses, are in principle falsifiable in Popper's sense, although the $M-M$ experiment is compatible with both \footnote{In 1934, Popper considered that the FitzGerald-Lorentz hypothesis had no falsifiable consequences, but
he later corrected himself and accepted Grünbaum's (1959) critique. See Popper (1992, p. 83 and note 1).}.  Nevertheless, a crucial experiment is conceivable that tests \footnote{It does not follow from this that the Kennedy-Thorndike experiment (1932), using an instrument with
arms of unequal lengths, is the crucial experiment to have refuted the contraction hypothesis, as is often
claimed (see a detailed discussion of this point in Laymon, 1980).}  them, for example, by accurately measuring the times taken by light rays to travel down each arm of the instrument and back. Such an experiment is physically possible, although technologically unfeasible. In any case, such an experiment could not be decisive since it would have its own presuppositions, which, in turn, could also be questioned. 

Let us note the conceptual differences between the Michelson and Morley and Lorentz interpretations of the same experimental data $\Delta N = 0$. For the former, the time that measures the duration of the phenomena is absolute and the length of a material body is invariant, that is, independent of its state of motion. Also, the speed of light is not constant but is combined with the speed of the Earth according to the Galilean transformations. This leads them to interpret that the unobserved difference in the velocity of light with respect to the movement of the Earth implies that the Earth does not move with respect to the medium in which the light is propagated, meaning, it drags the ether. For Lorentz, on the other hand, the ether is at absolute rest, which according to him means that none of its parts moves with respect to the others. Consequently, it cannot be dragged by the Earth. Since Lorentz shares the assumption that time is absolute and the speed of light on Earth is not constant, he considers that the best possible explanation is to assume that the length of bodies is relative to their motion with respect to the stationary ether. 

To be acceptable, both the ether drag hypothesis and the hypothesis of the length contraction of moving bodies needed to be justified beyond accommodating the null result of the $M-M$ experiment. It was also required to demonstrate that such hypotheses were compatible with other optical and electromagnetic phenomena established at that time, such as the aberration of starlight. Their authors, especially Lorentz, attempted to do so, but we will not follow the intricate explanations they offered here \footnote{It does not follow from this that the Kennedy-Thorndike experiment (1932), using an instrument with
arms of unequal lengths, is the crucial experiment to have refuted the contraction hypothesis, as is often
claimed (see a detailed discussion of this point in Laymon, 1980).}.  Instead, we can simply observe that these hypotheses were conservative since they aimed at preserving the existence of the ether while adhering to the classical mechanics and electromagnetic theory of the time. On the other hand, even though there were other ways of interpreting the $M-M$ experiment, each of them implied questioning assumptions that were far more deeply rooted.
	
\section{Einstein's postulates} \label{section5}

Einstein presented his theory of special relativity deductively, though not rigorously axiomatized. He derived it from two fundamental postulates or hypotheses: the \textit{principle of relativity} and the \textit{principle of the constancy of the speed of light}. At the start of his foundational 1905 work, he formulated them in the following way:
\begin{quote}
    1. In all coordinate systems in which the mechanical equations are valid, also the same electrodynamic and optical laws are valid.
    
2. In empty space light is always propagated with a definite velocity $v$ which is independent of the state of motion of the emitting body. (Einstein, 1905b, p. 891-892; CP 2a, p. 140)

\end{quote}

These postulates have been commented on innumerable occasions; so, we won't expound much about them.  \footnote{See Zahar (1989), Ch. 2, for a precise exposition of the development of Lorentz electrodynamics, based
on the existence of the ether.} The first postulate affirms the equivalency of all inertial frames of reference, that is, those that do not rotate and move with uniform motion relative to each other. This hypothesis was firmly established in the realm of mechanics dating back to Newton. Yet, in the domain of electromagnetism, its validity was incompatible with the hypothesis of a stationary ether as a privileged frame of reference like absolute space. By postulating the validity of the principle of relativity for all physical phenomena with no restriction, Einstein was able to formulate an electrodynamic theory that no longer needed to maintain the existence of the ether. Hence his famous affirmation according to which, \say{the introduction of a \say{light ether} will prove superfluous} (Einstein, 1905b, p. 892; CP 2a, p. 141).

	With regards to the propagation of light, the second postulate affirms a property that is common to all wave phenomena: the speed of the waves does not depend on the movement of the source with respect to the medium of propagation. It does not imply that the speed of the waves is the same in all frames of reference that move with uniform motion relative to the medium in which they propagate. This makes the second postulate apparently incompatible with the first. However, as is well known, Einstein's relativity of simultaneity reconciles them.
	The two postulates of special relativity led to the revolutionary conclusion that the speed of light in a vacuum is the same in every inertial frame of reference, irrespective of the motion of the source and the observer. In other words, the invariance of the speed of light in a vacuum is a direct consequence of Einstein's two postulates. The relativity of distances and temporal intervals follows from the invariance of the speed of light. In turn, the ether's special status is no longer tenable. The frame of reference provided by the stationary ether loses its privileged character, as the speed of light has the same value c in any inertial frame of reference, including the ether.
 
	In later writings, Einstein stated that the two postulates were not enough to rigorously deduce the Lorentz transformations (for example, Einstein, 2002; CP 7a, pp. 126-127). It was necessary to postulate the homogeneity and isotropy of space (or, more precisely, of space-time), implicit in the 1905 formulation. However, the invariance of the speed of light, which follows exclusively from the two postulates, is all that is needed, as we shall see, to provide an explanation of the null result of the $M-M$ experiment.
 
	Einstein's interpreters have endlessly argued about the context of discovery of these two postulates. In particular, the question has been raised as to whether Einstein knew about the $M-M$ experiment before 1905 and whether he took it into account in formulating special relativity. We are not going to take a position on this delicate historical problem; nonetheless, we can summarize the situation as follows.
 
	Einstein did not cite the $M-M$ experiment in his original 1905 paper nor, significantly, in his 1946 autobiography (Einstein, 1949). In all the places where he mentioned this experiment, from 1907 to the end of his life, he never claimed that it played a significant role in the genesis of special relativity. \footnote{For a detailed analysis of Einstein's 1905 paper see Torretti (1996), Ch. 3. On the history of the
principle of relativity, see Paty (1999).}  It is unlikely that before 1905 he would have read Michelson's original articles on the subject. It is generally accepted, however, that Einstein knew the result of the $M-M$ experiment from reading Lorentz's 1895 essay, where it is discussed  in detail. Einstein's knowledge of this work by Lorentz is attested by various independent statements he made.  \footnote{See Holton (1969) and Shankland (1963) for further details and quotations of relevant passages.}

In the numerous places in which he refers to the origin of the two postulates of special relativity, Einstein affirms, over and over again, and always in a very concise way, that he took the principle of relativity from Newtonian mechanics, extending it to electromagnetic phenomena, and the principle of the constancy of the speed of light from the electrodynamics of Maxwell and Lorentz.  \footnote{See Shankland (1973); Pais (1982), Ch. 6; Stachel (1982) and Patty (1993), Ch. 3.} Special relativity consisted in making these two postulates compatible by replacing the Galileo transformations with the Lorentz ones. On the other hand, when Einstein refers to the experimental bases of his postulates, he mentions other physical phenomena: Faraday's electromagnetic induction as support for the principle of relativity; Fizeau's (1851) experiment on the propagation of light in a water current and Airy's (1871) experiment on the aberration of starlight measured with a telescope filled with water as support for the principle of the constancy of the velocity of the light. After 1905, every time he cites the $M-M$ experiment, he considers it as evidence in favor of the principle of relativity but never of the constancy of the speed of light. It is therefore reasonable to assume that if this experiment played any role in the genesis of special relativity, it was only partial, since Einstein took it as one among other experiments that supported his postulates. It does not seem to have had any influence on the adoption of the postulate of constancy of the speed of light, despite the claims of many works on relativity. \footnote{For example, Einstein (1919), (1923), and (1949), among many others. The place where Einstein
discusses the $M-M$ experiment in most detail is his unpublished work from 1920, first published only in
2002 (Einstein, 2002; CP 7). On the other hand, the only place where he would explicitly state that he
knew of this experiment before the formulation of special relativity is in his 1922 lecture delivered in
Kyoto, translated only in 1982 (Einstein, 1982). This is an unreliable source because no German text of
the lecture has survived, only its Japanese transcription. The 1982 translation has been challenged as
erroneous precisely in the passage where it refers to the $M-M$ experiment (see Itagaki, 1999). In any case,
there is no guarantee that the preserved text is a faithful reproduction of Einstein's words. The Japanese
version of Einstein’s lecture, together with a new English translation of it, was published in 2012
(Einstein, 2012; CP 13).} So, in principle, at least, Einstein could have formulated the two postulates of special relativity exactly as he did in 1905 without having known about the $M-M$ experiment.  \footnote{This point has been clarified by Stachel (1982) after examining all the passages in Einstein's work prior
to 1922 in which the $M-M$ experiment is mentioned. Textbooks on special relativity sometimes consider it
as evidence in favor of the constancy of the speed of light postulate (for example, Smith, 1995, Ch. 2) and
sometimes as an experimental proof of the relativity postulate (for example, Shadowitz, 1988, Ch. 9). The
first time Einstein mentioned the $M-M$, in his 1907 review article on special relativity, he was quite explicit on the fact that, according to him, the experiment provided evidence in favor of the first postulate
of his theory, the relativity principle:
\begin{quote}
    Michelson and Morley’s experiment had actually shown that phenomena agree with the principle
of relativity even where this was not to be expected from the Lorentz theory. It seemed therefore
as if Lorentz’s theory should be abandoned and replaced by a theory whose foundations
correspond to the principle of relativity, because such a theory would readily predict the negative
result of the Michelson and Morley Experiment”. (Einstein 1907, p. 413; CP 2, p. 253).
\end{quote}
}

One of the reasons for the success of Einstein's theory was undoubtedly its axiomatic character. His two postulates are extremely simple, but they have remarkable consequences. Let us consider here only those that refer to the existence of the ether and to the invariance of the speed of light. 

	Above all, the first postulate does not imply that the ether does not exist, as is often said, but only that it is not necessary for the formulation of electrodynamics. Weyl, for example, in his great work on relativity states that:
 \begin{quote}
     The only reasonable answer that was given to the question as to why a translation in the ether 	cannot be distinguished from rest was that of Einstein, namely, that \textit{there is no ether!} (Weyl, 	1952, p. 172).
     
 \end{quote}
	
	Einstein, however, did not subscribe to this interpretation. In the original 1905 article, he only said that:
 
  \begin{quote}
	The introduction of a\say{light ether} will prove superfluous, inasmuch as in accordance with 	the concept to be developed here, no \say{space at absolute rest} 	endowed with special properties 	will be introduced, nor will a a velocity vector be assigned to a point of empty space at which 	electromagnetic processes are taking place. (Einstein, 1905b, p. 892; CP 2a, p. 141)
  \end{quote}
  
	Fifteen years later, Einstein recognized that in 1905 he had thought that \say{one should no longer talk about the ether in physics}, but that \say{this judgement was too radical} (Einstein, 2002; CP 7a, p. 130). On this point, he wrote that:
 
  \begin{quote}
	[…] It is still permissible to assume a space-filling medium whose states may be imagined as 	electromagnetic fields (and perhaps also as matter). But it is not permissible to attribute to these 	medium states of motion in every point, like in analogy to ponderable matter. This ether must not 	be imagined as consisting of particles whose identity could be traced in time”. (Einstein, 2002; 	CP 7a. p. 130) 
   \end{quote}
   
	After formulating the general theory of relativity, Einstein vindicated the existence of the ether since, according to this theory, space had physical properties. In a famous lecture delivered in 1920, he drew the following conclusion:

\begin{quote}
[...] According to the general theory of relativity, space is endowed with physical qualities, therefore; in this sense, therefore there exists an ether. According to the general theory of relativity space without ether is unthinkable; for in such space there not only would be no propagation of light, but also no possibility of existence for standards of space and time (measuring-rods and clocks), nor therefore any space-time intervals in the physical sense. But this ether may not be thought of as endowed with the quality characteristic of ponderable media, as consisting of parts of parts which may be tracked through time. the idea of motion may not be applied to it. (Einstein, 1920; CP 7a, pp. 181-182).
\end{quote}

	Consequently, the special relativity -locally valid in general relativity- is \textit{not incompatible} with the existence of the ether. It requires, however, that the idea of an ether at rest analogous to absolute space be abandoned. In Einstein’s terms, \say{we must by abstraction take from it the last mechanical characteristic which Lorentz had still left it} (Einstein, 1920; CP 7a, p. 171). This, in turn, implies that it is not possible to detect any motion with respect to the ether, since the ether is not in any state of motion. While experiments of the $M-M$ type continued to be performed within the framework of traditional ether theory until the 1930s, special relativity predicted a negative result for all of them, as indeed it happened.
 
	The invariance of the speed of light did not come, then, from an experimental result but constituted a theoretical hypothesis that allowed us to explain the null results of all the experiments of the $M-M$ type from a new perspective. Naturally, to maintain the constancy of the speed of light in all inertial frames of reference, it is necessary to deprive the lengths of bodies and the temporal duration of physical processes of an invariant character. These properties become relative to each inertial referential as expressed in the Lorentz transformations. Precisely, these transformations, according to Einstein's interpretation, lead to a new law for the addition of velocities,  a law that leaves the speed of light invariant. From this perspective a new mechanics emerges that replaces Newtonian mechanics.
 
	Einstein's originality consisted in postulating two basic hypotheses that had as consequences  the dissolution of the problems that were discussed in the context of the theory of the electromagnetic ether. The equivalence of all inertial frames of reference for the formulation of electrodynamics was evidently in conflict with the existence of a privileged referential such as the stationary ether. Before 1905, the validity of the relativity principle of mechanics was in doubt with respect to  electrodynamics; Einstein simply \textit{got rid of} the problem by $postulating$ unrestrictedly the validity of the relativity principle in the domain of mechanical, optical, and electromagnetic phenomena. However, an attempt could have been made to resolve the conflict between mechanics and electromagnetism by modifying the latter theory. Einstein's solution, consisting of maintaining the electromagnetic theory and modifying the mechanics, was much more radical and revolutionary than all the attempts to rearrange or modify the conceptions of the ether. In the long term, but not immediately, special relativity was much more successful than any of the research programs for electromagnetism based on the existence of that medium. Such programs persisted for many years after 1905 but gradually lost the consensus of the scientific community. The simplicity and unifying power of Einstein's theory made it so that special relativity was widely accepted, even at the cost of having to accept unintuitive consequences like the invariance of the speed of light.

\section{The interpretation of the experiment after 1905}\label{section6}

The explanation that the theory of special relativity provides for the result of the $M-M$ experiment is extremely simple. It is based on the invariance of the speed of light. Indeed, if the speed of light is the same in every inertial frame of reference, the $M-M$ experiment should give the same result whether the Earth moves around the Sun as if it were at rest with respect to it. In both cases, the total time spent by light in a round trip through each arm of the instrument when measured in a fixed referential to this instrument is $T_0 = 2L_0 / c$ (where $T_0$ is the proper time between the events of emission and reception of the light ray and $L_0$ is the proper length of each arm of the instrument, which are assumed to be equal). Since the speed of light is $not$ combined with the speed of the Earth, the rotation of the instrument arms has no effect on the speed of light and therefore on the observed interference fringes.

	It follows from  the relativistic law for the addition of velocities (which is a consequence of the invariance of the speed of light)  that, for any velocity $v$, $c + v = c - v = c$. If we apply this law to the $M-M$ experiment, we get the result that the time $T_1$ taken for a ray of light to travel back and forth along the arm oriented parallel to the Earth’s orbital velocity $v$ is the same as the time taken by that ray if the Earth were at rest relative to the Sun (or the cosmic ether), that is, $2L/c$. Using the Galilean transformation for velocities, as Michelson and Morley did, we obtain $T_1 = \frac{L}{c-v} + \frac{L}{c+v}$, which, as we have seen, equals $\frac{2L}{c(1-\frac{v^2}{c^2})}$ . Since the term $1-\frac{v^2}{c^2}$  is less than 1, $T_1$ is greater than $2L/c$. However, if we apply the relativistic transformation for velocities, it immediately results that $T_1 = 2L/c$. So, everything happens as if the Earth were at rest relative to the ether. Within the framework of special relativity, any attempt to measure the relative velocity of the Earth with respect to the ether must give a null result, because the theory predicts that, due to the invariance of the speed of light, the experiment must give the same result on a stationary or a moving Earth. 

According to special relativity, if an observer moves with velocity $v$ in the direction of one of the arms, the length of the arm would be shortened by a factor equal to $(1-v^2/c^2)^{1/2}$, while the arm length perpendicular to the direction of motion would remain unchanged. In such a case, the round-trip times taken by the light, as measured by the mobile observer, would not be the same. In the arm that is parallel to the movement of the observer, the time would be $ T_1 = 2L_0 (1-v^2/c^2)^{1/2} / c$, but in the arm that is perpendicular to the direction of the movement, the time would be $T_2 = 2L_0 / c$. It's evident that $T_1$ is less than $T_2$, thus the beams do not return simultaneously to the source. This is an example of the relativity of simultaneity. In special relativity there are only relative movements of the frames of reference (according to the first postulate); any experiment, mechanical or electromagnetic, must give the same result in all inertial frames of reference. Therefore, in principle, it is impossible to detect absolute motion, or absolute velocity, such as the $M-M$ experiment was intended to find. In this way, the null result is explained, not only of this experiment, but of all experiments of the same type. \footnote{The link between the $M-M$ experiment and the theory of special relativity seems to have been forged
early. By 1920 it was already common to present special relativity in textbooks following a historical
sequence that begins with the first attempts to detect the movement of the Earth with respect to the ether,
continues with the $M-M$ experiment and its different interpretations, and culminates with the theory of
Einstein. It is often suggested, though not always stated explicitly, that special relativity was proposed to
explain the outcome of the $M-M$ experiment more satisfactorily than Lorentz's theory did. This plot
scheme is followed by works such as Born (1962), originally published in 1920, and, to mention one
written in Spanish, Cabrera (1923). In greater or lesser detail, works such as Møller (1952), Bohm (1965),
French (1968), Resnick (1968), Sartori (1996) and most popular books adopt this classical approach.
Even works that do not take the historical approach almost always mention the $M-M$ experiment in
connection with Einstein's postulates, e.g. Weyl (1952), Bridgman (1965), Landau and Lifchitz (1994),
Ellis and Williams (2000) (where it is confused with later experiments carried out with starlight), Taylor
and Wheeler (1992) and Schutz (1993), among many others that could be cited. Einstein himself
presented a historical sequence similar to the classical approach in his unpublished paper from 1920,
published in 2002 (Einstein, 2002; CP 13). In his informative book of 1917, Einstein only mentions the
$M-M$ experiment at the end of his exposition of special relativity and presents it as one of the experiences
that confirm his theory (Einstein, 1917, §16). It is significant that in the Prologue to this work, Einstein
affirms that he exposes his ideas respecting, in general, the order and context in which they actually arose.}

	From the point of view of the context of the discovery of special relativity, it can be said that Einstein did not build this theory to explain supposed experimental evidence provided by the $M-M$ experiment or any other about the movement of the Earth in the ether. Einstein's primary objective was to solve a conceptual problem: the incompatibility between Newtonian mechanics and Maxwell's electrodynamics. However, in the context of justification, the $M-M$ experiment has operated and still operates as a confirmatory element of his theory. In fact, it constituted one of the first evidence for special relativity. It is not an example of  the prediction of a new fact, that is, unknown before the formulation of the theory, but  a case of accommodation of previously known evidence (the absence of shifting of the interference fringes). This result is interpreted in a completely new way: as a consequence  of the invariance of the speed of light, a phenomenon that was inconceivable when Michelson and Morley carried out their experiment. Since special relativity allows us to deduce the null result of the $M-M$ experiment, this evidence acts as a confirmatory element of the theory. Philosophers of science have extensively discussed the problem of the confirmatory value of this class of evidence, that is, whether the prediction of new phenomena is more or less important than the accommodation of known phenomena. However, there is no doubt that scientists consider that a theory is confirmed by the data it entails, especially when it is believed that (as in this case before 1905) there is no fully satisfactory alternative explanation of those data.   \footnote{On the relativistic analysis of the $M-M$ experiment from the point of view of a referential in movement
with respect to the interferometer, see Schumacher (1994). This case is too complex to examine here.}
 
    While the $M-M$ experiment confirmed special relativity, it was by no means sufficient to justify the acceptance of this theory. In general, the process that leads to the consensual acceptance of a theory and the rejection of its rivals requires a great deal of time. It is necessary to show that the new theory can satisfactorily reinterpret and explain other experimental results related to its domain of application. In addition, it is necessary to somehow discredit rival theories, for example by pointing out that they fail to explain certain experiments that are more satisfactorily explained by this new theory.

In fact, after 1905, other hypotheses were proposed that made it possible to account for the results of the M–M experiment in a different way from that of the theory of special relativity. One of these was the Ritz emission theory, formulated in 1908. According to Ritz, light can be conceived as being composed of point particles emitted in all directions by an electrical charge. The speed of these light particles is equal to c with respect to the source that emits them. However, this speed is not invariant. If the source moves with speed $v$ with respect to a referential $K$, the speed of light in $K$ will be equal to the vectorial sum of $v$ and $c$. Ritz's theory, as noted, uses the Galilean transformations to relate different inertial systems. This emission theory provides an immediate explanation for the results of any $M-M$ type experiment where the light source is terrestrial and at rest with respect to the measuring instruments. In these experiments, both the light source and the mirrors are at rest relative to the interferometer. Consequently, the speed of light is the same in all directions in the referential associated with the apparatus. The round-trip times of the light are equal to $2L/c$ in each arm of the instrument. According to both theories, the phenomena would occur in the same way if the Earth were at rest or in motion with respect to the ether. In this sense, Ritz's theory can also dispense with the ether as a privileged referential. 

And well, by 1910, the emission theories still provided a reasonable explanation for the negative result of the $M-M$ experiment. They also had the benefit of preserving classical mechanics and Galileo's transformations. Its weakness was that Maxwell and Lorentz's theory of electromagnetism had to be modified because the speed of light depended on the $motion$ of the source. Einstein himself admitted that sometime before 1905 he had considered the possibility of adopting a theory of emission, but soon abandoned it because he could think of \say{no form of differential equation which could have solutions representing waves whose velocity depended on motion of the source} (Shankland, 1963, p. 49). We note here Einstein's commitment to the wave theory of light as he thought exclusively in terms of a theory of wave emission. It is intriguing that, having postulated the light quantum hypothesis in the same year 1905 (actually, a few months before writing the paper on special relativity), he did not consider the possibility of a particle theory of emission.

Now, the theory of special relativity and the theory of Ritz emission were not, in fact, empirically equivalent since they predicted different results for the case in which the light source was in relative motion with respect to the Earth. Suppose that a source outside the Earth, such as the Sun, moves with velocity $v$ relative to the (terrestrial) measuring device. Special relativity predicts that in this case the time travel of the light in each arm of the instrument must be equal to $2L/c$ in the instrument's frame of reference since the speed of light is independent of the speed of the source. Ritz's theory, on the other hand, predicts that the time travel measured in the arm of the apparatus aligned with the direction of motion of the source will be equal to $L/(c+v) + L/(c-v) = 2L / c(1 – v^2/c^2)^{1/2}$, which agrees with the calculation of Michelson and Morley. Consequently, an experiment of the $M-M$ type carried out with a non-terrestrial light source would make it possible to discriminate between both theories. According to Ritz's theory, the experiment should give different results at different times of the year since the speed of light should vary due to the combination of the Earth's rotation and translation motions. Experiments by Tomaschek in 1924 using starlight and by Miller in 1925 using sunlight gave the same negative results as the $M-M$ experiment. These results have generally been considered unfavorable evidence for any emission theory.  \footnote{Glymour (1980) offers a classic discussion of this issue. A typical example of accommodation of
previously known evidence is the explanation offered by the theory of general relativity of the anomaly in
the advance of the perihelion of Mercury. There is no doubt that the scientific community considered it as
a confirmation of this theory. According to Brush (1989), it was even more valuable than the other two
classical tests, redshift and light bending. There is, however, an important difference regarding the role of
the $M-M$ experiment in special relativity. The theory of general relativity also had its origin in a
conceptual problem: the incompatibility between the Newtonian theory of gravitation and special
relativity, more precisely, the fact that Newton's universal law of gravitation was not covariant with
respect to the Lorentz transformation. This fact led Einstein to search for a relativistic theory of
gravitation. However, Einstein was well aware of the Mercury perihelion anomaly and took it into
account in his long search for the relativistic equations of the gravitational field. One of the conditions
these equations had to meet was precisely to explain this anomaly. A letter to Conrad Habicht, as early as
1907, says so explicitly:
\begin{quote}
    At the moment I am busy with considerations on the theory of relativity in connection with the
law of gravitation... I hope to clear up the hitherto unexplained secular changes of Mercury's
perihelion longitude... but so far it doesn't seem to work. (Quoted in Pais, 1982, p. 190).
\end{quote}}

\section{Conclusion: The experiment in its context}\label{section7}

Many experiments of the $M-M$ type continue to be performed to this day. In a certain way, they are part of the routine of contemporary science and are even carried out as laboratory practices in basic physics courses. The most sophisticated ones provide an increasing degree of precision. \footnote{The Ritz emission theory was a genuine rival theory to special relativity, a counterproposal, as
Sommerfeld called it (Preface to Pauli, 1958, p. xi). It had wide repercussions around the 1910s, being
discussed, among others, by De Sitter, Ehrenfest, and Tolman (for full references see Pauli, 1958, pp. 5-
8). By 1921, however, Pauli already considered it untenable because, although it explained the $M-M$
experiment, it could not account for other optical experiments, such as that of Fizeau (Pauli, 1958, p. 9).
After Tomaschek's (1924) and Miller's (1925) experiments with sunlight and stellar light, Ritz's theory
was considered disproved (see references in Miller, 1933 and Shankland et al., 1955). In 1965, however,
J. G. Fox made a detailed analysis of all the experimental evidence contrary to the emission theories and
showed that it was inconclusive before the meson experiments carried out in 1964 (Fox, 1965, p. 16).}  However, the theoretical context in which such experiments are interpreted has radically changed from the time of Michelson and Morley. They are no longer considered as attempts to measure the speed of the Earth with respect to the ether but as tests of the postulates of special relativity. As Swenson points out, physicists generally teach that the rise of relativity occurred after the fall of the ether, but historians must counter that the fall of the ether occurred after the rise of relativity (Swenson, 1970, p. 63). \footnote{The $M-M$ experiment was compatible with the existence of a 1/6 $v$ ether wind since this was the
precision threshold of their instruments. Subsequent experiments gradually lowered this threshold. In that
of Cedarholm \textit{et al }(1958), the maximum speed of the possible ether wind was reduced to 1/1000 $v$. In
experiments with laser light, such as that of Brillet and Hall (1979), or the more recent one by Müller et
al. (2003) this $v$ threshold was lowered by a factor of 4,000 and 12,000, respectively. It is evident that if
these experiments gave a positive result, it could not be attributed to an effect of the terrestrial movement
since it would be a tiny fraction of the orbital speed of the Earth.} In this context, the expectations have been reversed: the null results are what is expected, and each one is interpreted as an increasingly precise confirmation of special relativity. Any positive result would be surprising as it would contradict the predictions of one of the best-established theories in physics. At present, a difference of one part in a million in the measurements of the speed of light in different directions would very likely put the theory of special relativity in crisis since it would lead to a questioning of the invariance of the speed of light. This hypothesis did not arise, as we have seen, as a product of observations or experiments but, on the contrary, it was used to explain them. The \say{refutation} of this hypothesis would involve a drastic reinterpretation of many other experiments. 

We have seen the different interpretations that the M–M experiment received before and after 1905. If we now imagine a different historical context, an experiment of the $M-M$ type could have been interpreted as refuting at least some of the auxiliary hypotheses that are common to all post-1887 interpretations of this experiment. For this purpose, let us recall our discussion about the different kinds of assumptions that operate in the interpretation of an experience. 

The hypothesis of an orbital movement of the Earth, for example, was shared by all the interpretations that we have commented on and was beyond any reasonable doubt. Considering all that has been discussed, let us ask ourselves, then, what would have happened if an experiment of this kind had been carried out at a historical moment in which said hypothesis was not yet considered confirmed. The question is reasonable if we take into account the independent development that electromagnetism has had with respect to mechanics, which allows us to suppose that, in principle, an experiment of the $M-M$ type could have been carried out in the 17th century (that is, when the hypothesis of the movement of the Earth  was in full discussion) under experimental conditions analogous to those of the late 19th century. In other words, in principle, an experiment of this type could be carried out in the context of Huygens' wave theory of light and the null result of the experiment would have been interpreted, naturally, as confirmation of the hypothesis that the Earth was at absolute rest with respect to the luminiferous ether.
 
Already the measurement by Römer in 1676 of the speed of light considering the delay of the eclipses of the satellite Io when passing behind Jupiter, $assumed$ the orbital movement of the Earth. Indeed, the maximum delay in the initiation of the eclipse was supposed to be produced by the fact that the Earth and Jupiter were at a maximum distance, so that the light had to travel the diameter of the Earth's orbit employing the greatest additional time. However, the phenomenon was compatible with a system of the type of Tycho Brahe and even that of Ptolemy. Tycho Brahe did not observe the phenomenon since he did not have a telescope (the discovery of Io was announced by Galileo in 1610), but he could have explained it by attributing the retardation to the diameter of Jupiter's epicycle $centered$ on the Sun (which in turn orbited around the Earth). Regarding Ptolemy's system, it could have been assumed that Jupiter's epicycle was the size of the Earth's Copernican orbit. Of course, these auxiliary hypotheses would have required adjustments in the corresponding theories.

The first experience that was regarded as observational evidence of the orbital movement of the Earth was offered from the interpretation of the aberration of starlight by Bradley in 1728, a phenomenon that he attributed to the terrestrial movement. Let us remember that Bradley was looking for the parallax of a star to precisely measure the orbital speed of the Earth: he wanted to determine the apparent Keplerian ellipse described by a star throughout the year. But he found, on the one hand, that the shape in which the ellipse was drawn was not as expected in relation to the different speeds and directions of the Earth's movement. On the other hand, he found that all the stars described ellipses whose semi-major axis was independent of their distance. Bradley interpreted, then, that the effect was not due to the position of the star with respect to the Earth (distance, location on the ecliptic, etc.) but to its apparent speed; even more, the sought-after parallax was hidden (and it was only measured by Bessel more than a century later, in 1838). Could not this phenomenon have been interpreted assuming that all the stars were on the same vault and that it moved in an elliptical fashion? 

Let us note, in passing, the enormous weight acquired by the wave theory of light, especially from 1801 with the interference phenomenon studied by Young. Einstein solved the ether problem, upholding Maxwell's $wave$ electromagnetic theory to its ultimate consequences, in which the speed of a wave did not depend on the speed of the source. Thus, the aberration of light, difficult to interpret from the wave theory throughout the 19th century, was not enough to reveal, as the photoelectric effect did for Einstein in the same year 1905, that light could have corpuscular behavior. 

Finally, other possible situations could be imagined in the same historical context in which Michelson began his experiments. If the Lorentz contraction hypothesis had been formulated before Michelson's first interference experiment, for example, in the context of a discussion about whether the density of matter depended on its state of motion, the Lorentz theory could have been the hypothesis that was sought to be tested. It would have required some acceptable atomic theory. Consequently, later, the result of the $M-M$ experiment would have been interpreted within the framework of said atomic theory and would have been considered as confirmation of the hypothesis that the length of bodies contracted with movement. From this point of view, the aim of the experiment would not have been the determination of the absolute speed of the Earth with respect to the ether; still less, the measurement of the invariance of the speed of light.

 	Every experiment is carried out based on a very broad set of theoretical assumptions. Which of them are considered as the hypothesis that the experiment intends to test and which as mere auxiliary hypotheses that are not tested depends on the consensus reached at a given historical moment. In a different context, if the consensus were different, the experiment would have a completely different meaning. For example, the role of the auxiliary hypotheses could be exchanged so that one of the hypotheses, previously considered well-established, takes the place of the main hypothesis that is put to the test in the experiment. So much so that the same experiment -particularly that of Michelson and Morley- could, depending on the context, be considered astronomical, mechanical, electromagnetic, purely optical, or even atomic.
  
 	To understand the impact that Einstein's 1905 article had historically, we must contextualize it; that is, understand which auxiliary hypotheses were accepted and which phenomena were explained. Special relativity achieved a synthesis of different epistemic virtues. Minimizing the number of postulates -statements that, by definition, do not require explanation- and adopting some auxiliary hypotheses, he obtained, through few elements, an admirable explanatory power and simplicity. At the time Einstein formulated it, it was the only explanation of the $M-M$ experiment that was compatible with all known phenomena about the propagation of light, such as stellar aberration, Fizeau's experiment, and many others.  Also, unlike other theoretical alternatives, it maintained its empirical adequacy with respect to all optical and electromagnetic experiments carried out after 1905 and up to the present. For this reason, the acceptance of special relativity can be attributed to the confluence of three virtues: having been presented at the precise moment in which an apparently insoluble conflict between mechanics and electromagnetism was recognized; to have been formulated by simply $accepting$ as basic facts certain phenomena on the propagation of light that until then had been unsuccessfully tried to explain, and, finally, to have served as an explanatory framework for all known optical and electromagnetic phenomena.

    In principle, an experimental result is never definitively established. It is perfectly possible that in the future the $M-M$ experiment will be interpreted based on an a theory entirely different from special relativity and taken, for example, as having no relation to the invariance of the speed of light or to any of the Einstein postulates, but to some yet unknown property of light. The relativistic explanation employs, like any other, a broad set of assumptions that are taken for granted up to now but we cannot determine whether they will continue to be accepted in the future. Hacking has argued that the result of the $M-M$ experiment constitutes one of those \say{permanent facts about phenomena that any future theory must accommodate} (Hacking, 1983, p. 254). This was undoubtedly true for Lorentz, Einstein, and Ritz, and it still is today. However, for the reasons just stated, we cannot accept the independence of experimental facts from any interpretive theory. It is possible that in the future, in an entirely different theoretical context than the one accepted today, the result of the $M-M$ experiment will not be regarded as a fact that needs to be explained or accommodated, or even conceptualized as a fact at all.

$Acknowledgments$: We thank Rafael Ferraro and Michel Paty for their observations and comments on a previous version of this work.

\end{document}